# A TASK OF THE STORAGE CONTROL THEORY IN TRANSPORT SYSTEMS USING RESAMPLING-METHOD

**Helen Afanasyeva**
Faculty of  Computer science and Electronics,
Transport and Telecommunications Institute, 1 Lomonosov Str.,
LV-1019 Riga, Latvia, jelena_a@tsi.lv, +371 9186405

**ABSTRACT:** The probability of shortage absence is estimated for the storage of some transport system. The intensive computer methods of statistics are used in corresponding processes simulation. The efficiency of suggested approach, taking the mean square error of the estimator as the criterion, is illustrated with corresponding numerical examples.

## INTRODUCTION

The term storage in contemporary business finishes to be simply calculating indicator of activity, but becomes one of the most important objects of management, that influence the success of all transport institution. Taking into account growing mobility of nowadays transport it is very important to organize considering logistics processes of storage supply more efficiently. The specialists suggest to analyze and to optimize such complex stochastic systems on base of corresponding mathematical models and simulation.

Let us consider some transport company that has a number of microbuses. There is storage of very rare spares, that are not used very often, but their state of shortage is very expensive. What is the optimal level on this storage? One of the ways to solve such problem is to build the corresponding simulation model. If we have sufficient data of time moments of supply and demand we can estimate the corresponding probability laws and their parameters. Then we can use this information as the input flow into our simulation model of storage control.

The main problem that can arise here is that the input data sizes are not sufficiently big to predict the right probability law and estimate parameters correctly. The suggested nonparametric resampling-approach helps to avoid those disadvantages of traditional methods. In contemporary world with powerful modern computers were opened new perspectives before the statisticians. Intensive computer methods of statistics [5], or resampling methods allows to use the input data for simulation without previous estimation of distribution lows or their parameter simply in different combinations. This makes possible to turn the data with the different sides and to get the target parameters of simulation without bias and with smaller value of variance. Resampling gives robust estimators of parameter of interest, taking as efficiency criterion the mean square error.

In the present paper the resampling-approach implementation is considered to a task of storage control theory [1] in transport systems.

Suppose we have two simple independent renewal processes: demand $\{X_i,\ i=1,2,...\}$ and supply $\{Y_i,\ i=1,2,...\}$, where $\{X_i\}$ and $\{Y_i\}$ are the sequences of unnegative independent random variables, each sequence with its own common distribution [6],[3]. The distribution functions of sequences $\{X_i\}$ and $\{Y_i\}$ are unknown, but corresponding initial samples of sizes $n_X$ and $n_Y$ are available. The renewals of the processes correspond to the supply or to the demand of storage units. The initial storage level $K$ is known. Then the probability of interest $P\{D_m>S_{m-K}\}$, of the shortage absence, is the probability, that the $m$-th demand $D_m=\sum_{i=1}^{m} X_i$ comes later, than the $m-K$-th supply $S_{m-K}=\sum_{i=1}^{m-K} Y_i$ .

Then the probability of interest, of the shortage absence, is the probability, that the $m$-th demand comes later, that the $m-K$-th supply $D_m>S_{m-K}$. It is also assumed, that the initial storage level $K$ is known. We wish to investigate some properties of the different estimators of the shortage absence probability.

In the next section some important relationships for future purposes are given. Then follows the section, where the resampling-estimators of probability of interest are presented. The third section presents the specific cases for some distributions on the base of resampling-approach. The fourth section considers the classical estimators of the probability of interest. Further in the fifth section numerical examples illustrate the suggested approach efficiency. The last section concludes the paper.

## 1. SOME RELATIONSHIPS

Now we describe our problem more formally. Let us define the distribution functions of sequences $\{X_i\}$ and $\{Y_i\}$ as $F_1^X(x)$ and $F_1^Y(x)$, and distribution density functions as $f_1^X(x)$ and $f_1^Y(x)$. The functions $F_1^X(x)$ and $F_1^Y(x)$ are unknown, but corresponding samples $H^X=\{X_1,X_2,\ldots,X_{n_X}\}$ and $H^Y=\{Y_1,Y_2,\ldots,Y_{n_Y}\}$ are available for sequences $\{X_i\}$ and $\{Y_i\}$, where $|H^X|=n_X$ and $|H^Y|=n_Y$.

Let us define the distribution and density function of the sum of variables: $F_i^X(x)$ and $f_i^X(x)$. The sub index $i$ in this notation means the number of addends in the sum. The mentioned above function $F_1^X(x)$ according to this notation is the distribution function of the sum presented with only one element. The upper index means the r.v. name, whose distribution function is considered.

We are interested in the time of the $m$-th and $m-K$-th renewals:

$$D_m=\sum_{i=1}^{m} X_i\ ,\ S_{m-K}=\sum_{i=1}^{m-K} Y_i\ . \qquad (1)$$

Our task is to estimate the shortage absence probability $P\{D_m>S_{m-K}\}$ that the $m$-th renewal of the demand process $\{X_i\}$ comes later, than the $m-K$-th renewal of the supply process $\{Y_i\}$.

Let's consider the indicator function $\Psi(\mathbf{x},\mathbf{y})$, where $\mathbf{x}=(x_1,x_2,\ldots,x_{m_X})$ and $\mathbf{y}=(y_1,y_2,\ldots,y_{m_Y})$ are vectors of real numbers:

$$\Psi(\mathbf{x},\mathbf{y})=\begin{cases}1 & if \sum_{i=1}^{m_X} x_i > \sum_{i=1}^{m_Y} y_i \\ 0 & otherwise.\end{cases} \quad (2)$$

Suppose we have two vectors of r.v. $\mathbf{X}=(X_1, X_2, \ldots X_{m_X})$ and $\mathbf{Y}=(Y_1, Y_2, \ldots Y_{m_Y})$, $m_X = m$, $m_Y = m-K$. Our purpose is to estimate the shortage absence probability $\Theta = E(\Psi(\mathbf{X},\mathbf{Y}))$. We will estimate $\Theta$ using two different approaches: classical and resampling. The classical, parametrical approach is widely known. So we consider the alternative nonparametrical resampling-approach implementation in this paper.

## 2. RESAMPLING-APPROACH

At first we consider the resampling-approach for the estimation of the shortage absence probability. This method does not supposes the estimation of the distribution parameters or the construction of the empirical distribution functions to find characteristics of interest, as it is supposed by traditional methods. Alternatively we use primary data in different combinations and this fact makes possible to obtain unbiased estimators and decrease their variance. There are a lot of examples of resampling-approach successive implementation to various tasks for reliability systems, regression models, for order statistics estimation [2], [7].

Resampling-approach supposes the following steps. We choose randomly $m_X$ elements from the sample $H^X$ and $m_Y$ elements from the sample $H^Y$. The elements are taken without replacement, we remind that $n_X \geq 2m_X$, $n_Y \geq 2m_Y$. Then we calculate the corresponding value of function $\Psi(\mathbf{x},\mathbf{y})$ using formula (2). After that we return chosen elements into the corresponding samples.

We repeat this procedure during $r$ realizations. Let $j_d^i(l)$, $d$=1,..,$m_i$ be the indices of elements from the sample $H^i$, $i \in \{X,Y\}$, that are chosen at the $l$-th realization. Then for the $l$-th realization we obtain the following vectors:

$$\mathbf{X}(l)=(X_{j_1^X(l)}, X_{j_2^X(l)}, \ldots, X_{j_{m_X}^X(l)}),\ \mathbf{Y}(l)=(Y_{j_1^Y(l)}, Y_{j_2^Y(l)}, \ldots, Y_{j_{m_Y}^Y(l)}).$$

The resampling-estimator $\Theta^{R*}$ is the arithmetical mean by $r$ realizations:

$$\Theta^{R*} = \frac{1}{r}\sum_{l=1}^{r}\Psi(\mathbf{X}(l),\mathbf{Y}(l)). \quad (3)$$

Obviously this estimator is unbiased:

$$E(\Theta^{R*}) = \Theta. \quad (4)$$

We are interested in the variance of this estimator.

Let's denote the following notations:

$$\mu = E(\Psi(\mathbf{X},\mathbf{Y})),\ \mu_2 = E(\Psi(\mathbf{X},\mathbf{Y})^2), \quad (5)$$

$$\mu_{11} = E\big(\Psi(\mathbf{X}(l),\mathbf{Y}(l))\Psi(\mathbf{X}(l'),\mathbf{Y}(l'))\big), l \neq l'. \quad (6)$$

Then the variance of interest can be calculated using the following formula:

$$V(\Theta^{R*}) = E(\Theta^{R*2}) - \mu^2, \tag{7}$$

where

$$E(\Theta^{*2}) = \frac{1}{r}\mu_2 + \frac{r-1}{r}\mu_{11}. \tag{8}$$

In order to estimate the variance of the estimator, we have firstly to find the expression of the mixed moment $\mu_{11}$ from the formula (6).

To calculate the moment $\mu_{11}$ the notation of α-pairs can be used [4]. Let us denote $W_i(l)$, $l=1,...,r$, $i \in \{X,Y\}$, a subset of the sample $H^i$, which was used for producing the values of vectors $\mathbf{X}(l)$ and $\mathbf{Y}(l)$ correspondingly, $W_i(l) \subset H^i$. Let us denote $M_i=\{0,1,...,m_i\}$, $M=M_X \times M_Y$. Let $\boldsymbol{\alpha}=(\alpha_X, \alpha_Y)$ be an element of $M$, $\boldsymbol{\alpha} \in M$. We say that $W_i(l)$ and $W_i(l')$ produce the $\boldsymbol{\alpha}$-pair, if and only if $W_i(l)$ and $W_i(l')$ have $\alpha_i$ common elements: $|W_i(l) \cap W_i(l)| = \alpha_i$.

Let $A_{ll'}(\boldsymbol{\alpha})$ denote the event "subsamples $(\mathbf{X}(l),\mathbf{Y}(l))$ and $(\mathbf{X}(l'),\mathbf{Y}(l'))$ produce $\boldsymbol{\alpha}$-pair", but $P_{ll'}(\boldsymbol{\alpha})$ be the probability of this event: $P_{ll'}(\boldsymbol{\alpha})=P\{A_{ll'}(\boldsymbol{\alpha})\}$. Because of the fact realizations $l=1,..,r$ are statistically equivalent, we can omit the lower indices $ll'$ and write $P(\boldsymbol{\alpha})$.

Let

$$\mu_{11}(\boldsymbol{\alpha}) = E\big(\Psi(\mathbf{X}(l),\mathbf{Y}(l))\Psi(\mathbf{X}(l'),\mathbf{Y}(l')) \mid A_{ll'}(\boldsymbol{\alpha})\big), \tag{9}$$

then

$$\mu_{11} = \sum_{\boldsymbol{\alpha} \in M} P(\boldsymbol{\alpha})\mu_{11}(\boldsymbol{\alpha}). \tag{10}$$

Therefore we need to calculate $P(\boldsymbol{\alpha})$ and $\mu_{11}(\boldsymbol{\alpha})$ for all $\boldsymbol{\alpha} \in M$.

The probability $P(\boldsymbol{\alpha})$ can be calculated using hypergeometrical distribution**:**

$$P(\boldsymbol{\alpha}) = \prod_{i\in\{X,Y\}} \binom{m_i}{\alpha_i}\binom{n_i - m_i}{m_i - \alpha_i} \Big/ \binom{n_i}{m_i}, \tag{11}$$

where $\binom{n}{m}$ is binomial coefficient.

Now our task is to calculate $\mu_{11}(\boldsymbol{\alpha})$, $\forall \boldsymbol{\alpha} \in M$.

Let's introduce some new notations for two different realizations $l$ and $l'$ of the resampling-procedure. Using sums, mentioned earlier in formula (1) let's include the upper index corresponding to the realization number. Then let's divide each sum into two parts in the following way. We consider separately the common and the different elements of these sums for realizations $l$ and $l'$:

$$\begin{aligned} D^{l}_{m_X} &= D^{dif(ll')}_{m_X-\alpha_X} + D^{com(ll')}_{\alpha_X}, \quad D^{l'}_{m_X} = D^{dif(l'l)}_{m_X-\alpha_X} + D^{com(ll')}_{\alpha_X}, \\ S^{l}_{m_Y} &= S^{dif(ll')}_{m_Y-\alpha_Y} + S^{com(ll')}_{\alpha_Y}, \quad S^{l'}_{m_Y} = S^{dif(l'l)}_{m_Y-\alpha_Y} + S^{com(ll')}_{\alpha_Y} \end{aligned} \tag{12}$$

where $D^{j}_{m_X} = \sum_{\xi \in W_X(j)} X_\xi$ (or $S^{j}_{m_Y} = \sum_{\xi \in W_Y(j)} Y_\xi$) is the sum of sequences $\{X_i\}$ (or $\{Y_i\}$) elements for the $j$-th realization, $j \in \{l,l'\}$,

$D^{dif(jk)}_{n}$ (or $S^{dif(jk)}_{n}$) is the sum of $n$ elements from $\boldsymbol{X}(j)$ (or $\boldsymbol{Y}(j)$), which are absent in $\boldsymbol{X}(k)$ (or $\mathbf{Y}(k)$), $k, j \in \{l,l'\}, k \neq j$,

$D^{com(ll')}_{n}$ (or $S^{com(ll')}_{n}$) is the sum of $n$ common elements of $\mathbf{X}(l)$ and $\boldsymbol{X}(l')$ (or $\boldsymbol{Y}(l)$ and $\boldsymbol{Y}(l')$).

Therefore we can write:

$$\begin{aligned}\mu_{11}(\boldsymbol{\alpha}) = {} & P\{\Psi(\mathbf{X}(l),\mathbf{Y}(l))=1,\Psi(\mathbf{X}(l'),\mathbf{Y}(l'))=1\,|\,\boldsymbol{\alpha}\} = \\ = {} & P\{D^{dif(ll')}_{m_X-\alpha_X}+D^{com(ll')}_{\alpha_X} > S^{dif(ll')}_{m_Y-\alpha_Y}+S^{com(ll')}_{\alpha_Y}, D^{dif(l'l)}_{m_X-\alpha_X}+D^{com(ll')}_{\alpha_X} > S^{dif(l'l)}_{m_Y-\alpha_Y}+S^{com(ll')}_{\alpha_Y}\} = \\ = {} & P\{D^{dif(ll')}_{m_X-\alpha_X}+C_{\boldsymbol{\alpha}} > S^{dif(ll')}_{m_Y-\alpha_Y}, D^{dif(l'l)}_{m_X-\alpha_X}+C_{\boldsymbol{\alpha}} > S^{dif(l'l)}_{m_Y-\alpha_Y}\} = \\ = {} & \int_{-\infty}^{+\infty} P\{D^{dif(ll')}_{m_X-\alpha_X}+z > S^{dif(ll')}_{m_Y-\alpha_Y}, D^{dif(l'l)}_{m_X-\alpha_X}+z > S^{dif(l'l)}_{m_Y-\alpha_Y}\} f^C(z\,|\,\boldsymbol{\alpha})dz,\end{aligned}$$

where $D^{com(ll')}_{\alpha_X} - S^{com(ll')}_{\alpha_Y} = C_{\boldsymbol{\alpha}}$, $f^C(x\,|\,\boldsymbol{\alpha})$ – the distribution density function of r.v. $C_\alpha$. Note that for fixed value of r.v. $C_{\boldsymbol{\alpha}} = z$ the events $\left\{D^{dif(ll')}_{m_X-\alpha_X}+z > S^{dif(ll')}_{m_Y-\alpha_Y}\right\}$ and $\left\{D^{dif(l'l)}_{m_X-\alpha_X}+z > S^{dif(l'l)}_{m_Y-\alpha_Y}\right\}$ are independent. Then $\mu_{11}(\boldsymbol{\alpha})$ has the following form:

$$\mu_{11}(\boldsymbol{\alpha}) = \int_{-\infty}^{+\infty} P\{D^{dif(ll')}_{m_X-\alpha_X}+z > S^{dif(ll')}_{m_Y-\alpha_Y}\}\cdot P\{D^{dif(l'l)}_{m_X-\alpha_X}+z > S^{dif(l'l)}_{m_Y-\alpha_Y}\} f^C(z\,|\,\boldsymbol{\alpha})dz\,.$$

Therefore

$$R(z\,|\,\boldsymbol{\alpha}) = P\left\{D^{dif(ll')}_{m_X-\alpha_X}+z > S^{dif(ll')}_{m_Y-\alpha_Y}\right\} = P\left\{D^{dif(l'l)}_{m_X-\alpha_X}+z > S^{dif(l'l)}_{m_Y-\alpha_Y}\right\} = \int_{-\infty}^{+\infty} F^Y_{m_Y-\alpha_Y}(x+z) f^X_{m_X-\alpha_X}(x)dx. \qquad (13)$$

The r.v. $C_{\boldsymbol{\alpha}}$, has the following cumulative distribution and probability density functions:

$$\begin{gathered}F^C(z\,|\,\boldsymbol{\alpha}) = P\{C_{\boldsymbol{\alpha}} \le z\} = P\{D^{com(ll')}_{\alpha_X} - S^{com(ll')}_{\alpha_Y} \le z\} = P\{D^{com(ll')}_{\alpha_X} \le S^{com(ll')}_{\alpha_Y} + z\}, \\ f^C(z\,|\,\boldsymbol{\alpha}) = \int_{-\infty}^{+\infty} f^X_{\alpha_X}(x+z) f^Y_{\alpha_Y}(x)dx.\end{gathered} \qquad (14)$$

Note that the events $\left\{D^{dif(ll')}_{m_X-\alpha_X}+z > S^{dif(ll')}_{m_Y-\alpha_Y}\right\}$ and $\left\{D^{dif(l'l)}_{m_X-\alpha_X}+z > S^{dif(l'l)}_{m_Y-\alpha_Y}\right\}$ are equal probable. Therefore we can write the formula for $\mu_{11}(\boldsymbol{\alpha})$ calculation in the following way:

$$\mu_{11}(\boldsymbol{\alpha}) = \int_{-\infty}^{+\infty} R(z\,|\,\boldsymbol{\alpha})^2 f^C(z\,|\,\boldsymbol{\alpha})dz. \qquad (15)$$

## 3. SPECIFIC CASES

This section includes some examples, when we illustrate the suggested approach on well known distributions.

*Example: Exponential distribution*

Let's consider now another example, when in our renewal processes r.v. $X$ and $Y$ of interest have exponential distribution with parameters $\lambda$ and $\nu$ correspondingly. Then the distribution function $F^X_{m_X-\alpha_X}(x)$ of the sum $D^{dif(ij)}_{m_X-\alpha_X}$ from formula (12) has Erlang distribution with parameters $\lambda$ and $m_X - \alpha_X$. The distribution function $F^Y_{m_Y-\alpha_Y}(x)$ of the other sum $S^{dif(ij)}_{m_Y-\alpha_Y}$ from formula (12) has also Erlang distribution with parameters $\nu$ and $m_Y - \alpha_Y$. We also define with letter *G* (with corresponding indices*)* an additional function to corresponding distribution function: $G(x)=1-F(x)$.

Let us consider the integral from (13) in the following way:

$$R(c\,|\,\boldsymbol{\alpha}) = 1 - \int_{-\infty}^{+\infty} G^{Y}_{m_Y-\alpha_Y}(y+c) f^{X}_{m_X-\alpha_X}(y)dy = 1 - \int_{-\infty}^{\max(0,-c)} T(y\,|\,\boldsymbol{\alpha})dy - \int_{\max(0,-c)}^{+\infty} T(y\,|\,\boldsymbol{\alpha})dy,$$

where $T(y|\boldsymbol{\alpha})$ is the underintegral expression.

All intermediate proofs and calculus for those two integral parts are given in the [1]. Finally, we have:

$$R(c\,|\,\boldsymbol{\alpha}) = 1 - F^{X}_{m_X-\alpha_X}(\max(0,-c)) - e^{-\nu c}\lambda^{m_X-\alpha_X} \sum_{i=0}^{m_Y-\alpha_Y-1} \frac{\nu^i}{i!} \sum_{p=0}^{i} \binom{i}{p} c^p \frac{1}{(\lambda+\nu)^{m_X-\alpha_X-p+i}} \cdot$$
$$\cdot \frac{(i-p+m_X-\alpha_X-1)!}{(m_X-\alpha_X-1)!} \cdot G_X(\max(0,-c)\,|\,\boldsymbol{\alpha}), \quad (17)$$

where $G_X(x\,|\,\boldsymbol{\alpha})$ - additional function for Erlang distribution function with parameters $\lambda+\nu$, $m_X-\alpha_X+i-p$.

Now consider the probability density function from (14) for the r.v. $D^{com(ij)}_{\alpha_X} - S^{com(ij)}_{\alpha_Y} = C_{\boldsymbol{\alpha}}$, where $D^{com(ij)}_{\alpha_X}$ and $S^{com(ij)}_{\alpha_Y}$ have Erlang distribution with parameters ($\alpha_X$, $\lambda$) and ($\alpha_Y$, $\nu$) correspondingly.

Then

$$f^C(c\,|\,\boldsymbol{\alpha}) = \frac{\lambda^{\alpha_X}\nu^{\alpha_Y}e^{-\lambda c}}{(\alpha_X-1)!(\alpha_Y-1)!} \cdot \sum_{p=0}^{\alpha_X-1} \binom{\alpha_X-1}{p} \cdot c^p \frac{(\alpha_X+\alpha_Y-p-2)!}{(\lambda+\nu)^{\alpha_X+\alpha_Y-p-1}} G_X(\max(0,-c)), \quad (18)$$

where $G_X(x)$ corresponds to additional function for Erlang distribution with parameters $\lambda+\nu$ and $\alpha_X+\alpha_Y-p-1$. The proof for formula (18) are given in the [1]. Then using formula (15) we can find the necessary mixed moment.

## 4. CLASSICAL APPROACH

The classical approach to the estimation of the probability of interest is a parametrical one. It supposes the point estimation of the parameters of the distribution, if we know the distribution type of the initial samples $H^i$, $i=\{X, Y\}$. We intend to estimate the parameters of the known types of distributions.

*Example: Exponential distribution*

Let's consider an example, when r.v. *X* and *Y* have exponential distribution with parameters $\lambda$ and $\nu$ correspondingly. The sum of exponentially distributed r.v. has Erlang distribution. The probability of interest $\Theta = P\{D_{m_X} > S_{m_Y}\}$, notation from formula (1), now is:

$$\Theta = \int_0^{+\infty} G^X_{m_X}(y) f^Y_{m_Y}(y)dy = \int_0^{+\infty} e^{-\lambda y} \sum_{i=1}^{m_X} \frac{(\lambda y)^{i-1}}{(i-1)!} \nu \frac{(\nu y)^{m_Y-1}}{(m_Y-1)!} e^{-\nu y} dy = \sum_{i=0}^{m_X-1} \frac{\nu^{m_Y}}{(\lambda+\nu)^{m_Y+i}} \frac{\lambda^i}{i!} \prod_{p=0}^{i-1}(m_Y+p). \quad (19)$$

The classical approach supposes using the point estimators instead of the values of $\lambda$ and $\nu$: $\lambda^* = n_X / D_{n_X}$ , $\nu^* = n_Y / S_{n_X}$ . It gives the estimator:

$$\Theta^{C*} = \sum_{i=0}^{m_X-1} \frac{\nu^{*m_Y}}{(\lambda^* + \nu^*)^{m_Y+i}} \frac{\lambda^{*i}}{i!} \prod_{p=0}^{i-1} (m_Y + p), \qquad (20)$$

where $\prod_{p=0}^{-1} = 1$.

Now the expression for variance of $\Theta^{C*}$ is: $V(\Theta^{C*}) = E(\Theta^{C*2}) - E(\Theta^{C*})^2$ and for the mean square error $ER(\Theta^{C*}) = V(\Theta^{C*}) + (\Theta - E(\Theta^{C*}))^2$.

## 5. NUMERICAL RESULTS

Consider the case when r.v. $X$ and $Y$ have exponential distribution with parameters $\lambda$=0.3, $\nu$=0.7. The real estimators of probabilities of the shortage absence are presented in the Table 1 (the first row of each section).

Table 1

Experimental results for real probabilities of shortage absence $\Theta$, Classical $\Theta^{C*}$ and Resampling $\Theta^{R*}$ estimators of $\Theta$

| | | $K$=0 | $K$=1 | $K$=2 | $K$=3 |
|---|---|---|---|---|---|
| $n_1$=14 | $\Theta$ | .9218 | .9527 | .9747 | .9887 |
| $n_2$=13 | $B(\Theta^{C*})$ | .0862 | .0654 | .044 | .0249 |
| $m$=6 | $V(\Theta^{C*})$ | .0956 | .0711 | .047 | .0262 |
| $\lambda$=0.3 | $ER(\Theta^{C*})$ | .103 | .0753 | .0489 | .0269 |
| $\nu$=0.7 | $V(\Theta^{R*})$ | .0123 | .0085 | .0093 | .0081 |
| $n_1$=14 | $\Theta$ | .874 | .9295 | .9695 | .9919 |
| $n_2$=13 | $B(\Theta^{C*})$ | .1096 | .0808 | .0473 | .0167 |
| $m$=6 | $V(\Theta^{C*})$ | .1318 | .0963 | .0555 | .0195 |
| $\lambda$=0.5 | $ER(\Theta^{C*})$ | .1438 | .1028 | .0577 | .0198 |
| $\nu$=0.7 | $V(\Theta^{R*})$ | .0142 | .0237 | .021 | --- |
| $n_1$=10 | $\Theta$ | .7155 | .8046 | .8809 | .9391 |
| $n_2$=9 | $B(\Theta^{C*})$ | .1124 | .1066 | .0891 | .0624 |
| $m$=4 | $V(\Theta^{C*})$ | .1431 | .1292 | .1037 | .0701 |
| $\lambda$=0.3 | $ER(\Theta^{C*})$ | .1558 | .1406 | .1116 | .074 |
| $\nu$=0.7 | $V(\Theta^{R*})$ | .0551 | .0453 | .0482 | .0474 |

Let our sample sizes be equal $n_X$ and $n_Y$, and we consider the $m$-th unit's demand and different initial storage levels $K$=0..3. All calculations have performed for $r$ = 1000 realizations in the case of the resampling-approach.

We intend to compare the variance of the estimators of the resampling-approach with the mean square error of the classical approach. It is so because of the resampling-approach estimators are unbiased, but the classical ones on the contrary have bias.

In the Table 1 we can see the resampling-estimators' variance $V(\Theta^{R^*})$ comparing with classical approach estimators' variance $V(\Theta^{C^*})$, bias $B(\Theta^{C^*})$, and mean square error $ER(\Theta^{C^*})$.

The table shows how changes the results depending on different sample sizes *n*, unit's number *m* and initial storage level *K*.

Analyzing table's results we can draw the conclusion that the variance and corresponding mean square error of both approaches decreases with the increasing of sample sizes *n*, *m*, and initial storage level *K*. The variance of the resampling-estimators is almost always smaller than traditional one. However the resampling-estimators are unbiased. Taking as the criterion the mean square error resampling gives even better results for big values of *K*.

## CONCLUSION

The resampling-approach can be successfully used for obtaining the estimators of parameters of interest of the renewal processes. Obtained formulas allow calculating the variance of the estimators for the resampling and classical approaches. Numerical examples show the efficiency of suggested approach, taking estimators' mean square error as efficiency criterion. This approach can be the good alternative to traditional one.

## ACKNOWLEDGEMENTS

I would like to thank my scientific supervisor professor Alexander Andronov for his useful, valuable ideas and comments during this paper preparation.